%% file: manuscript.tex
\documentclass[manuscript]{acmart}

\usepackage[inkscapelatex=false]{svg}
\usepackage{float}
\usepackage{bm}
\usepackage{graphicx}
\usepackage[capitalize, nameinlink]{cleveref}
\usepackage{makecell}
\usepackage{listings}
\newcommand{\zenodoUrl}{\url{https://doi.org/10.5281/zenodo.19665253}}
\AtBeginDocument{%
  }

\setcopyright{rightsretained}
\copyrightyear{2026}
\acmYear{2026}
\acmDOI{XXXXXXX.XXXXXXX}
\acmConference[ICER '26 Vol. 1]{ACM Conference on International Computing Education Research V.1}{August 11--14, 2026}{Uppsala, Sweden}
\acmISBN{978-1-4503-XXXX-X/2026/08}
\begin{document}

\title[Fast and Forgettable]{Fast and Forgettable: A Controlled Study of Novices’ Performance, Learning, Workload, and Emotion in AI-Assisted and Human Pair Programming Paradigms}

\author{Nicholas Gardella}
\email{njg4ne@virginia.edu}
\orcid{0009-0000-6912-7462}
\affiliation{%
  \institution{University of Virginia}
  \city{Charlottesville}
  \state{Virginia}
  \country{USA}
}

\author{James Prather}
\email{jrp09a@acu.edu}
\orcid{0000-0003-2807-6042}
\affiliation{%
  \institution{Abilene Christian University}
  \city{Abilene}
  \state{Texas}
  \country{USA}
}

\author{Juho Leinonen}
\email{juho.2.leinonen@aalto.fi}
\orcid{0000-0001-6829-9449}
\affiliation{%
  \institution{Aalto University}
  \city{Espoo}
  \country{Finland}
}

\author{Paul Denny}
\email{p.denny@auckland.ac.nz}
\orcid{0000-0002-5150-9806}
\affiliation{%
  \institution{University of Auckland}
  \city{Auckland}
  \country{New Zealand}
}

\author{Raymond Pettit}
\email{rp6zr@virginia.edu}
\orcid{0000-0001-9675-025X}
\affiliation{%
  \institution{University of Virginia}
  \city{Charlottesville}
  \state{Virginia}
  \country{USA}
}

\author{Sara L. Riggs}
\email{sriggs@virginia.edu}
\orcid{0000-0002-0112-9469}
\affiliation{%
  \institution{University of Virginia}
  \city{Charlottesville}
  \state{Virginia}
  \country{USA}
}

\renewcommand{\shortauthors}{Gardella et al.}

\begin{abstract}
Code-generating Artificial Intelligence has gained popularity within both professional and educational programming settings over the past several years. While research and pedagogy are beginning to cope with this change, computing students are left to bear the unforeseen consequences of AI amidst a dearth of empirical evidence about its effects. Though pair programming between students is well studied and known to be beneficial to self-efficacy and academic achievement, it remains underutilized and further threatened by the proposition that AI can replace a human programming partner. In this paper, we present a controlled pair programming study with 22 participants who wrote Python code under time pressure in teams of two and individually with GitHub Copilot for 20 minutes each. They were incentivized by bonus compensation to balance performance with understanding and were retested individually on the programming tasks after a retention interval of one week. Subjective measures of workload and emotion as well as objective measures of performance and learning (retest performance) were collected. Results showed that participants performed significantly better with GitHub Copilot than their human teammate, and several dimensions of their workload were significantly reduced. However, the emotional effect of the human teammate was significantly more positive and arousing as compared to working with Copilot. Furthermore, there was a nonsignificant absolute retest performance reduction in the AI condition and a larger retest performance decrement in the AI condition. We recommend that educators strongly consider revisiting pair programming as an educational tool in addition to embracing modern AI.

\end{abstract}




\received{20 April 2026}
\received[revised]{N/A}
\received[accepted]{N/A}

\maketitle

\section{Introduction}
Generative Artificial Intelligence (GenAI) powers a number of tools programmers and computing students can use to assist in writing code, solving problems, and completing other coding related tasks \cite{denny_computing_2024}. These include general-purpose chatbots like ChatGPT and editor-based coding toolsets like GitHub Copilot \cite{bird_taking_2023}, the focus of this work. Copilot is branded as an "AI pair programmer," a nod to a team programming methodology popularized in the early 2000s called "pair programming" \cite{beck_extreme_2004, friedman_introducing_2021}. Because AI-powered behaviors can resemble human behaviors, GenAI is often framed as a human-like assistant that can help with work otherwise done alone or with another human. Though research has begun to compare the new AI-assisted paradigm to traditional solo and pair programming paradigms, the vast breadth of potential harms and benefits still remains largely unexplored~\cite{prather_beyond_2025}.

AI assistance poses special threats to computing students. A burgeoning ecosystem of educational AI might potentially help with learning, but commonplace tools developed for professionals (i.e., GitHub Copilot) may not~\cite{prather_beyond_2025,prather_its_2024}. In general, code-generating AI improves basic programming performance, but programmers can become reliant on its help \cite{jeswani_premature_2025, becker_programming_2023}. Then, when AI becomes unavailable or makes mistakes, programmers may be less practiced at the very manual intervention or critical-thinking called for most by such a situation. This is a studied phenomenon in human-automation interaction called an "irony of automation" and has many parallels to human-AI interaction \cite{bainbridge_ironies_1983}. 
Aside from over-reliance, pair programming may offer more than Copilot from a social perspective. Although AI is highly reliable and accurate as a programming assistant and presents a considerable advantage for computing students struggling to learn basic programming concepts, it falls short regarding its modality, which is currently limited to text and speech-to-text interaction. Human interactions, by contrast, typically involve facial expressions, body language, two-way verbal communication, etc. \cite{kuttal_trade-offs_2021}. That is, though Copilot may match or exceed the informational capacity of peer or instructor assistance, it could fail, for other reasons, to aid the learning and academic performance of a computing student as much as a human \cite{komocsi_academic_2026}.

In this study, we build upon recent work \cite{valovy_psychological_2023,lyu_will_2025, penney_understanding_2025, fan_impact_2025} by comparing human-human pair programming to programming with GitHub Copilot through the lens of the Control-Value theory of emotion. We conducted a mixed-methods study comparing performance, learning, workload, and emotion between these paradigms using a within-subjects design involving two study sessions one week apart. We found that despite lower workload and higher performance with GitHub Copilot, participants learned comparably and experienced more intense, positive emotions with their human partners. Our findings suggest the unique advantages of human-human programming over AI-assisted programming in computing education. We are guided by the following research questions:

\begin{itemize}
    \item \textbf{RQ1}: To what extent does actual and perceived \textbf{programming performance} differ between the human-human and human-AI team paradigms during a learning effort?
    \item \textbf{RQ2}: To what extent does \textbf{learning retention} from a programming effort differ between the human-human and human-AI team paradigms?
    \item \textbf{RQ3}: To what extent does the \textbf{workload} (i.e., mental, temporal, effort, frustration) associated with educational programming differ between the human-human and human-AI team paradigms?
    \item \textbf{RQ4}: To what extent does the \textbf{emotional impact} of educational programming as measured by emotional valence and arousal differ between the human-human and human-AI team paradigms?
\end{itemize}

\section{Related Work}

\subsection{Code-generating AI in Computing Education}

Empirical research about the educational impacts of code-generating AI assistants like GitHub Copilot is mixed. Though a number of tools \cite{kazemitabaar_codeaid_2024, liffiton_codehelp_2023, denny_prompt_2024} have been developed specifically for use in computing education, here we focus on the broader landscape of commodity GenAI such as Copilot and ChatGPT. AI assistants are frequently observed to enhance novices' programming performance \cite{fan_impact_2025, gardella_performance_2024, kazemitabaar_studying_2023}, but have yet to show similar effects on learning \cite{kazemitabaar_studying_2023, jost_impact_2024, xue_does_2024}. When computing students are allowed to use AI, they sometimes use it adaptively as a learning resource but other times use it to avoid active engagement with the material \cite{amoozadeh_student-ai_2024, margulieux_self-regulation_2024, prather_its_2024, prather_widening_2024, kazemitabaar_how_2024}. Avoidant behaviors of this nature are associated with worse learning in computing courses \cite{jost_impact_2024} and worse psychological well-being more generally \cite{dijkstra_engaging_2016}. Though AI can reduce workload \cite{hart_development_1988, gardella_performance_2024} and programming anxiety \cite{fan_impact_2025}, it may lack the conflict inherent within effective Socratic learning \cite{basawapatna_zones_2013, jonsson_cracking_2022}. Limited research about learning suggests that AI assistance is probably best used conservatively, in combination with traditional solo and paired paradigms \cite{kazemitabaar_how_2024, lyu_will_2025}. A possible avenue for enhancing learning would be increased motivation or self-efficacy, which tends to predict academic success \cite{honicke_influence_2016}. One study of 45 students found that using AI in a computing course significantly improved self-efficacy \cite{yilmaz_effect_2023}, and at least two studies have found benefits for motivation \cite{yilmaz_effect_2023, fan_impact_2025}. On the other hand, another study found that early and frequent AI use in a computing course was most common among students with low self-efficacy and course grades \cite{margulieux_self-regulation_2024}. Though one study found that programmers take credit for AI's successes and blame it for failures \cite{valovy_psychological_2023}, multiple others found the opposite \cite{gardella_performance_2024, hou_effects_2024, nam_using_2024, weisz_perfection_2021}. Overall, it seems that use of commodity AI in computing education brings much needed help at the cost of autonomy. Code-generating AI substantially changes the programming process \cite{barke_grounded_2023}, leading students to lose metacognitive control and awareness of appropriate behaviors \cite{shoufan_can_2023, prather_widening_2024}.

\subsection{Pair Programming in Computing Education}

Pair programming is highly regarded in computing education for its positive effects on student performance and well-being. Multiple meta-analyses have found that programming in pairs allows novices to work faster and tackle more challenging problems than they could alone, with lower or comparable effort \cite{umapathy_meta-analysis_2017, hannay_effectiveness_2009, hawlitschek_empirical_2023}. Beck et al. argued for the productivity of pair programming for professionals, too, encouraging it as a tenet of the Extreme Programming (XP) method to "write all production programs with two people sitting at one machine" \cite{beck_extreme_2004}. Pairing can even be more productive than a divide-and-conquer approach because of collaborative synergy. In the context of education, pair programming is generally more enjoyable and better for learning than solo programming \cite{umapathy_meta-analysis_2017, hanks_pair_2011}. Students also tend to believe in its merits, reinforcing them by a placebo-like effect \cite{hanks_pair_2011, colagiuri_can_2011}. Challenges with educational pair programming usually stem from issues of partner compatibility (either expertise or personality) and logistical concerns. Though it is typically best to pair students of similar ability level, diversity of perspectives may stimulate more productive discussion \cite{hanks_pair_2011, salleh_empirical_2011}.

\subsection{Learning in Human-Human v. Human-AI Pairs}

To reconcile potential benefits of AI and human teammates, several studies have compared the two paradigms experimentally \cite{kuttal_trade-offs_2021, valovy_psychological_2023, lyu_will_2025, penney_understanding_2025, fan_impact_2025}. In 2021, before modern GenAI, Kuttal et al. compared a human tutor to a simulated automated agent scripted and performed by a human in the wizard-of-oz style \cite{kuttal_trade-offs_2021}. They found that multi-modal communication and more lengthy, in-depth discussions were advantages of the human tutor. A later study interviewed software engineers and programming students after they programmed alone, in pairs, and with AI (ChatGPT and Copilot) \cite{valovy_psychological_2023}. They found participants were motivated by the alleviation of mundane tasks and satisfied by AI assistance, but preferred human teammates for their better knowledge, energy, and scrutiny.   

Moving into education, Penney et al. conducted a between-subjects qualitative experiment of 20 CS1 students who studied a concept through text chat with either an AI tool or a human tutor \cite{penney_understanding_2025}. The students generally found human assistants more trustworthy and reliable for learning but liked AI assistants better for fast, low-stakes, judgment-free exploration. The authors argue that novices see AI help-seeking as low-risk and low-reward, while they see human tutoring as high-risk and high-reward. Lyu et al. conducted a within-subjects study in an advanced web development course \cite{lyu_will_2025}. 39 students completed programming assignments in human-human, human-AI, and human-human-AI teams. Assignment scores were highest in the human-human-AI condition, followed by human-human and then human-AI. Qualitative data showed weaker partners were substantially helped by their stronger counterparts, but sometimes stronger partners felt held back. Pairs of weak partners struggled to make progress together; AI helped with this, but also dominated the team in the process. Participants felt that they could communicate ineffectively and still obtain helpful results. This marks a potential harm to the metacognitive skill of analyzing and expressing a question or goal, which has previously been considered as helped by the act of prompting \cite{prather_its_2024}. Finally, Fan et al. reported a quasi-experimental between-subjects study of 234 students in a web development course who programmed in either solo, human-human, or human-AI paradigms \cite{fan_impact_2025}. They found that performance was comparably higher in both paired conditions compared to the solo condition. Programming anxiety was most reduced by the human-AI paradigm, but perceived collaboration and social presence was highest in human-human teams.

\subsection{Theoretical Frameworks: Control-Value and Cognitive Load}

%Our theoretical framework draws from Control-Value Theory to consider the difference between human-human and human-AI programming teams through the important lens of emotional affect.
Our theoretical framework draws from Pekrun's Control-Value Theory (CVT) \cite{pekrun_control-value_2006, pekrun_control-value_2024} to consider the difference between human-human and human-AI programming teams through the important lens of emotional affect.
%Some early concepts of emotion in educational psychology described it to be negative in valence and variable by physiological arousal, with high arousal being bad (i.e., anxiety). For example, Bandura’s Social Cognitive Theory (SCT), which is popular in computing education research \cite{batra_emotions_2025}, argues that high \textit{self-efficacy}---one's belief in their own ability to succeed---motivates adaptive behaviors that improve academic performance and decrease the negative arousal associated with failure \cite{bandura_self-efficacy_1977}. A more useful theory for comparing human-human and human-AI paradigms is Pekrun's Control-Value Theory (CVT). 
CVT describes how humans' emotions follow from appraisals of both control over an achievement context and the value of achievement outcomes. When learning activities are low in value, students can feel low-energy negative emotions like boredom, for example, while feeling in-control during studying for a high value grade can induce strong positive emotions like excitement. 
%As in Russell's circumplex model of affect \cite{russell_circumplex_1980}, Roseman's Appraisal Theory of emotion \cite{roseman_appraisal_1996}, and others like them, CVT conceptualizes emotion with at least two basic dimensions: valence (positive v. negative) and arousal (low v. high energy). 
As with related emotional theories \cite{russell_circumplex_1980,roseman_appraisal_1996}, CVT conceptualizes emotion with at least two basic dimensions: valence (positive v. negative) and arousal (low v. high energy). 
A third dimension in CVT is the object focus, which has been expanded recently beyond only the achievement activity and its outcome to also include epistemic incongruities (informational conflicts), social beings (self and others), and existential feelings of vitality (health and mortality). In the context of educational team programming, relevant foci are the activity (programming itself), the outcomes (satisfaction, grades, or money), incongruities (differences in teammates' perspectives), the self (a focus for pride or shame), and others (foci for liking or commiseration) \cite{pekrun_control-value_2024}. Any of these foci can be the object of any given combination of valence and arousal. Research in writing studies and the neuroscience of creativity likewise shows that motivation and persistence in cognitively demanding work are mediated by limbic-system activation, especially in social or emotionally meaningful contexts \cite{flaherty_midnight_2004}. Human collaborators naturally stimulate the emotional and motivational circuits that sustain effort (see also \cite{sommers_revision_1980, sommers_responding_1982, vygotsky_mind_1978, paul_extended_2021}), while AI tools—pleasant but affectively flat—rarely trigger the same engagement. 

Considering emotional foci is useful because it broadens the possibilities for how to reach productive and enjoyable emotional states. If only considering outcome value, an AI assistant that helps complete satisfying projects or earn key grades might be chosen over a human teammate for promoting student well-being. AI assistance can also disrupt metacognition, leading students to feel overconfident in their own programming abilities on account of AI-driven progress \cite{prather_widening_2024}. The programming activity itself might feel more pleasant with AI for this reason. Similarly, lessened incongruities might make AI seem more emotionally preferable, since its design necessarily tries to inspire contentment, surprise, and delight over disagreement and conflict \cite{cheng_sycophantic_2025}. Though in isolation undesirable, conflict can be a catalyst not only for learning \cite{jost_impact_2024}, but also for coping \cite{roseman_appraisal_1996} skills that promote emotional resilience \cite{dijkstra_engaging_2016}. Unfortunately, sycophantic AI appeals to students' avoidant tendencies that seek to acclaim the righteousness of their existing perspectives \cite{cheng_sycophantic_2025}. We hypothesize that conflict, whether good or bad, is actually an essential adaptive property of human pair programming that AI is not currently effective at mimicking. Furthermore, AI's excessive competence compared to novices' is a dangerous precursor for it to assume control over programming tasks \cite{prather_widening_2024, lyu_will_2025}, flattening the emotional reward of a positive outcome value according to CVT. This could also lead to a focal transition to the self. Though Bandura's theory associates high self-efficacy to low arousal, we fear that low arousal and neutral to negative valence are harmful to pride and thus self-efficacy \cite{pekrun_control-value_2024}. Kallia labels this phenomenon as "AI Obscurity," whereby the credit a student normally enjoys for themself is given instead to the AI assisting them \cite{kallia_be_2025}. Finally, for social foci, it is unclear how the emotional impacts of pseudo-social interactions with AI compare to those of true human social interactions. Once again, sycophantic AI creates a hazardous attraction to seek the safe agreement of a supportive AI tool, abandoning rewarding human interactions in the process \cite{kallia_be_2025, cheng_sycophantic_2025, gardella_hbcu_2025}.

% self-regulated learning (SRL) theory??
% Theory of Planned Behavior \cite{ajzen_theory_1991} was used by Skripchuk \cite{skripchuk_investigation_2024} to explain how students seek help from AI as compared to web search. 

Finally, we consider the related phenomenon of workload through the lens of Cognitive Load Theory (CLT) \cite{sweller_cognitive_2011}. CLT states that learning activities have \textit{intrinsic} difficulty as standalone tasks as well as \textit{germane} difficulty from extracting generalizable skills and knowledge from the experience. Finally, \textit{extraneous} load adds superfluous difficulty to the learning process due to secondary minutia such as, in the case of programming, unclear instructions or confusing interfaces. With a limited cognitive capacity, students experience that these three sources of load stack, meaning excessive extraneous load leaves too little capacity left to effectively learn from the activity. Regarding extraneous load, qualitative research is mixed regarding AI's effect for programmers. While AI offloads considerable effort for expressing the syntax of a particular problem-solving approach \cite{ross_programmers_2023,kazemitabaar_how_2024,liang_large-scale_2024, gian_luca_scoccia_exploring_2023}, interacting with it adds new sources of confusion and overload \cite{barke_grounded_2023, prather_its_2024}. More importantly, there is no accepted standard for an optimal level of total workload on learning tasks, though research from disparate fields suggests that some moderate balance is best \cite{akizuki_measurement_2015, basawapatna_zones_2013}. Clearly, if AI removes an activity's intrinsic load, there is no stimulus to encourage growth  \cite{basawapatna_zones_2013}. Therefore, drawing from CVT, we anticipate human-human programming to bring together higher workload with strong positive emotions, in line with activity-based, social, and conflict-centric emotional object foci rather than an outcome focus.

\section{Method}

\subsection{Participants} \label{sec:participants}

The study was conducted in accordance with the Declaration of Helsinki and approved by the Institutional Review Board (IRB) for Social and Behavioral Sciences (SBS) at the University of Virginia (IRB-SBS \#7557) on July 22, 2025. Electronic informed consent was obtained from each participant. The IRB permitted us to mislead and deceive participants to support the study design, so we formally debriefed participants about this deception at the conclusion of the study. They were allowed to withdraw their consent at any time; no participants were concerned about the deception or withdrew consent.

Twenty-two participants were recruited by convenience sampling via email advertisements to roughly 900 engineering and data science students and in-class announcements to roughly 800 undergraduates across two sections of 2000-level CS (CS2; Data Structures and Algorithms) and three sections of 3000-level CS (two of Software Development and one of CS3; Data Structures and Algorithms II). Eligibility criteria included minimum age of 18, English fluency, novice or intermediate Python programming ability, and willingness to adhere to the study's multi-stage protocol. This protocol included a requirement to abstain from studying Python-related material for one week between study sessions, precluding most CS1 students from participating. Of 62 prospective participants, 25 followed through to a screening for honesty about their ability level (see screening tasks in our online appendices\footnote{\zenodoUrl}). Two were found ineligible for sub-novice ability levels, one dropped out after screening, and none were found ineligible for super-intermediate ability level. The remaining 22 participants completed the full study.

The participants were mostly male (16 male; 6 female) and Asian (15 Asian; 10 white; 1 Hispanic/Latinx) and were aged 18-29 ($\bar{x}=20.8$; $s=3.7$; $\tilde{x}= 18$). For ethical reasons, they were not asked where they learned about the study; however, the median age of 18 suggests that many were undergraduates in their first post-secondary year. Regarding disabilities and impairments, one participant self-reported having a physical impairment, and three reported having a cognitive or attentional impairment; none were excluded from the study. All participants had at least some experience with general purpose AI chat bots, while all but five had at least some prior experience with GitHub Copilot or a similar tool. As another measure of programming ability, we administered the Steinhorst \cite{steinhorst_revisiting_2020} self-efficacy instrument. Summary statistics ($\bar{x}=5.6$; $s=0.8$; $\tilde{x}= 5.6$) indicated that participants, on average, had self-efficacy consistent with students from other institutions at the end of their introductory (CS1) courses, i.e. \cite{steinhorst_revisiting_2020} ($\bar{x}=5.5$; $s=1.0$).% \cref{fig:steinhorst-hist} shows the distribution of self-efficacy scores for all participants.

% Systems Engineering - 400
% Data Science- 200
% Link Lab- 350
% 2100- 132+148
% 3140- 210+145
% 3100- 136

\subsection{Setup} \label{sec:setup}
The programming setup included two Windows desktop computers, two identical 32" monitors, two keyboards, and two mice. Two versions of the Visual Studio Code (VS Code; v1.104.1) integrated development environment (IDE) were configured on Windows 11 with Python 3.11.13 and the Pytest testing framework. Both had the standard Python extension, but one also had the GitHub Copilot (v1.372.0) extension with Chat (v0.31.1). The Chat extension was configured to use GPT 4.1---the flagship unlimited model included with Copilot's educational free tier at the time. The non-Copilot IDE had standard Python IntelliSense auto-complete from the Python extension. Two experimental workspace folders\footnote{See online supplement: \zenodoUrl} and one training folder contained sets of HumanEval tasks consisting of one PDF prompt, one Python file with an empty function to implement, and one testing file. The IDE was set up to run and test code either graphically or in the terminal. Participants were permitted to use the Google Chrome web browser and the Google search engine. AI search overviews were disabled, and using any AI in the browser was not allowed. Browsing history was tracked to ensure adherence to this restriction.

\subsection{Tasks}
Tasks were selected from a subset of the HumanEval dataset\footnote{\url{https://github.com/openai/human-eval}}~\cite{chen_evaluating_2021} used for a prior study \cite{gardella_performance_2024}. These hand-written Python tasks are often used to benchmark LLMs and AI code generators \cite{yetistiren_assessing_2022,naveen_comparative_2025}; each task comes with an English prompt and a set of automated tests. The prior study involved CS1 students completing timed HumanEval tasks either alone or with the assistance of Copilot. Eight tasks, each with a completion time of roughly 5 minutes, were organized into two pools of four tasks as similar as possible in difficulty to match the planned 20-minute time limit. 
%\cref{fig:expected-task-durations} shows the expected completion time distributions of the task pools under the treatment and control conditions of the prior study. 
The HumanEval tasks used were: Pool X (14, 52, 72, 0) and Pool Y (35, 9, 24, 18). 
Based on the prior study, the expected completion times of the tasks should be very similar across the two task pools. In the prior study, students completed the tasks in approximately 10 minutes on average with AI assistance, and in approximately 25 minutes solo. 
As mentioned in \cref{sec:setup}, task prompts were encoded in a PDF format. The skeleton code was stripped of the prompt text so that Copilot could not ingest it automatically. Instead, any context beyond the function stub had to be provided by the participant. Copilot was known to be very reliable for the tasks at hand, as this is the case for many simple educational programming tasks \cite{finnie-ansley_my_2023}. We confirmed that Copilot Chat was able to generate a passing solution for each of the eight tasks in one shot by copying the text of the PDF into the chat and asking Copilot to solve the problem. Participants were not advised of this or any other strategy, but were aware that Copilot could not read the contents of the PDF on its own.

\subsection{Experimental Design}
\begin{figure}
\vspace{-12pt}
    \centering
% \includesvg[width=\linewidth]{visuals//method/counterbalancing-scheme-cropped-top}
\includegraphics[width=\linewidth]{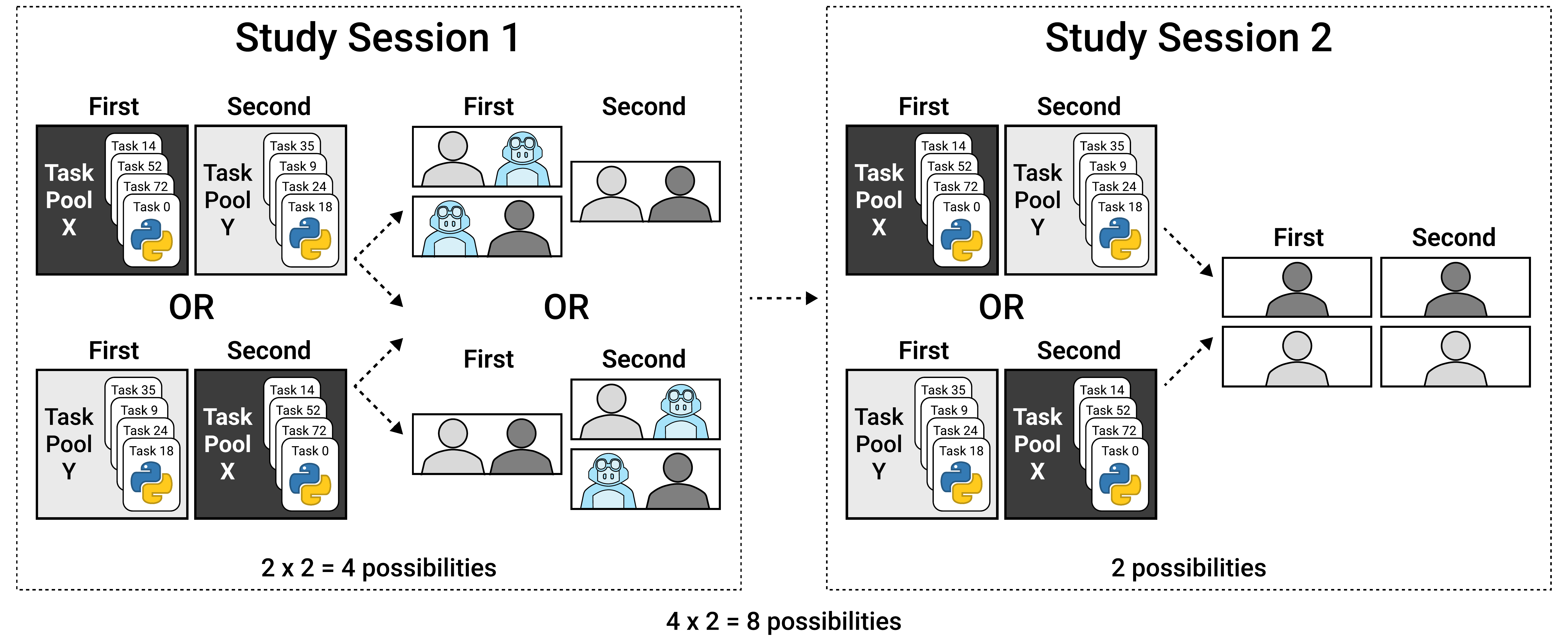}
    \caption{Experimental design and counterbalancing permutations.}
    \label{fig:design}
\end{figure}

The study used a within-subjects design with the within-subjects factor of team type (Human-Human v. Human-AI). For the first study session, two participants worked together once and separately with Copilot once. Roughly one week later, they each repeated both trials separately (without Copilot). \cref{fig:design} shows the eight design possibilities that were counterbalanced (rotated) across participants.

\subsection{Procedure}
As a prerequisite for study eligibility, participants were required to attend a free one-hour workshop about Github Copilot, its benefits and drawbacks, and its major features. Next, they had to pass a screening to demonstrate their attention during the workshop and confirm their ability level as novice or intermediate by attempting the screening tasks (see our online appendices\footnote{\zenodoUrl}) using a think-aloud technique in the presence of an experimenter. Eligible participants were paired based on their schedules and the requirement to have no prior relationships between programming partners.

\begin{figure}
    \centering
    % \includesvg[width=\linewidth]{visuals//method/procedure}
    \includegraphics[width=\linewidth]{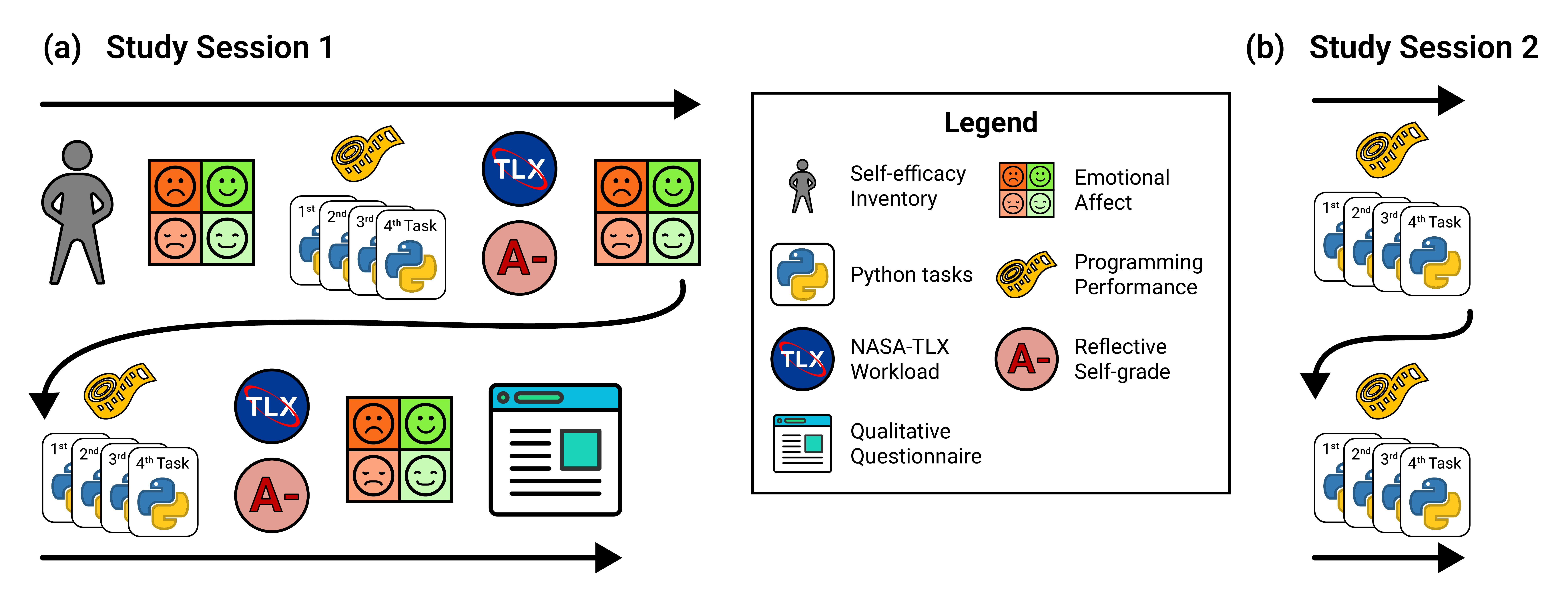}
    \caption{Procedure---from left to right, (a) shows the procedure for study Session 1, the first exposure to the programming tasks (in Human-Human or Human-AI teams), and (b) shows the procedure for study Session 2, the second exposure to the programming tasks (solo retention test).}
    \label{fig:procedure}
     \Description{A two session procedure is shown. Session 1 has a self-efficacy inventory, followed by two programming trials with performance measures, surrounded by emotion measures, followed by reflective self-grades and workload inventories, and ended by a single qualitative questionnaire. Session 2 has only two programming trials with performance measures.}
\end{figure}

On the day of their first session, a pair of participants met in the lab. They provided electronic informed consent and watched an orientation and training video\footnote{\url{https://www.youtube.com/removed-for-review}}. Their assigned goal was to "balance \underline{productivity} and \underline{understanding} the code" as they programmed. They were deceived to believe that a final score would be calculated by a secret weighting between their Session 1 score (productivity) and their Session 2 "retention test" (understanding) score (see \cref{eq:scoring}). They were also deceived to believe that they needed a top 15\% score to win a \$10 bonus on top of their base compensation. In reality, there was no grand score calculated and all participants were given the bonus. Furthermore, information about the nature of the "retention test" was withheld from the participants until the second week, when they were informed that it was simply a repeat of the tasks from the prior week. Following the training video, participants demonstrated their competencies on a practice task pool consisting of the three screening tasks. They practiced running and testing code and using Copilot before demonstrating their competencies and understanding of the rules. They also completed a short background questionnaire that contained the Steinhorst self-efficacy inventory \cite{steinhorst_revisiting_2020}. Next, they completed two 20-minute trials (one in each condition) according to one of the four possible design permutations (see \cref{fig:design}) along with questionnaires about emotion, workload, self-perceived performance, and their qualitative experience (see timeline in \cref{fig:procedure}a). Given they were balancing performance and understanding, they were instructed to use any extra time to study their solutions with their teammate, i.e., if they completed all four tasks in less than 20 minutes. Prior to the human-human trial, they were briefed on the popular driver and navigator pair programming model. They were not required to use this model and were allowed time to discuss their teamwork strategy before the timer started. At the end of each trial, code was saved in its final state for post-hoc scoring against the unit tests from HumanEval.

Following a roughly one week gap that varied slightly based on scheduling availability ($\bar{x}=\tilde{x}=1.0$, $x_{\min}=0.8$, $x_{max}=1.29$), participants returned for Session 2 where they took the retention test consisting of repeating each of the task pools from the prior week alone (see \cref{fig:procedure}b). Before leaving, participants were formally debriefed about the deception involved in the study in accordance with the approved IRB protocol. Compensation of \$7 per half hour plus a flat \$10 bonus was sent to each participant as a digital gift card.

% \begin{center}
%   \begin{minipage}[c]{0.52\textwidth}
\begin{equation}
  \begin{aligned}
    \text{Score}_{\text{ total}} = (\text{Score}_{\text{ team}}) \cdot (?) + (\text{Score}_{\text{ solo retest}}) \cdot (1 - ?)
  \end{aligned}
  \label{eq:scoring}
\end{equation}
% \end{minipage}
% \hfill
% \begin{minipage}[c]{0.40\textwidth}
\begin{equation}
  \begin{aligned}
    \text{Score}_{\text{ trial}} = \frac{20\ \text{pts}}{\text{task}} + \frac{5\ \text{pts}}{\text{task}} \cdot (\%\ \text{time to spare})
  \end{aligned}
  \label{eq:score}
\end{equation}
%   \end{minipage}
% \end{center}

% \begin{equation}
%     \text{Final Score} = (\text{Score Today}) \cdot (?) + (\text{Retention Test Score}) \cdot (1 - ?)
%     \label{eq:scoring}
% \end{equation}

\subsection{Data Collection}
\label{sec:datacollection}
\subsubsection{Emotion}
% \textbf{Emotion.} 
Participants reported their subjective emotional valence and arousal three times during Session 1 (see \cref{fig:procedure}a) according to Russell's \cite{russell_circumplex_1980} circumplex model of affect. They used seven-point Likert scales (coded 1-7). 
%accompanied by a visual aid similar to \cref{fig:emo-vis}. 
To account for emotional interference between trials, the dependent measures were changes in valence and arousal ratings (after - before).

\subsubsection{Workload}
% \textbf{Workload.} 
A digital version 
%(see \cref{fig:tlx}) 
of the popular six-item NASA Task Load Index (NASA-TLX) was used to measure workload, the toll that performance takes on a human \cite{hart_development_1988}. Since TLX asks about workload over a period of time, it is not represented as a difference from a baseline like emotion.

\subsubsection{Performance and Learning}
% \textbf{Actual Performance.} %\label{sec:measures-perf}
Session 1 team performance was quantified on a 100-point scale by a function (see \cref{eq:score}) 
that combines the number of tasks completed (out of four, worth 20 points for each task) with the amount of time remaining (out of the 20 minutes allocated) at the moment that the task count was first reached. Learning was quantified via Session 2 solo performance scores using the same formula.
%of the number of tasks completed out of four and the amount of time remaining on the clock when that score was first achieved. 
This approach provides internally valid ratio scaling properties; i.e. completing more tasks and completing them more quickly are rewarded proportionally,
%scoring twice as high or twice as fast is always rewarded appropriately, 
and completing fewer tasks very fast can never be rewarded more than completing more tasks (e.g., completing two tasks instantly would score 50, whereas completing three tasks with no time to spare would score 60).

\subsubsection{Perceived Performance}
%\label{sec:self-grade}
We also collected a self-grade from participants, as we were not confident the NASA-TLX performance dimension distinguished clearly between one's own performance and that of their teammate.

\subsubsection{Qualitative Data}
% \textbf{Qualitative Data.}
At the very end of Session 1, participants were asked 
%(see \cref{sec:exit-quest}) 
to type free-response text comparing and contrasting the two teaming conditions and describing in what ways each was advantageous over the other. The exact questions can be found in our online appendices\footnote{\zenodoUrl}.

\subsection{Data Analysis}

All analyses were conducted in R \cite{ihaka_r_1996} with a significance threshold of $\alpha=.05$. To estimate the effect of the treatment (AI teammate) relative to the control (Human teammate), we used Ordinary Least Squares (OLS) regression with Studentized Wild Cluster Bootstrapping (WCB) for inference \cite{webb_reworking_2023, mackinnon_fast_2023, cameron_bootstrap-based_2008}.  WCB was chosen to address the small sample size and dependence in the data, including repeated observations per participant and nesting within human pairs (i.e., each participant belongs to exactly one team, so observations from teammates are correlated). Compared to alternatives, this approach provides reliable inference under multi-way clustering with few clusters \cite{cameron_practitioners_2015, webb_reworking_2023}.

We tested for potential confounds due to task order or difference in the two task pools by performing Wald tests using \lstinline[language=R]{lmtest::waldtest(vcov = sandwich::vcovBS(R = 999, type = "wild-webb"))} \cite{zeileis_diagnostic_2002, zeileis_various_2020, zeileis_econometric_2004}. These factors did not significantly improve model fit ($p > .1$ in all cases), indicating that the study design and counterbalancing was effective; therefore, single-predictor models were used for all research questions.

Final inference was performed using \lstinline[language=R]{fwildclusterboot} using \lstinline[language=R]{boottest(B = 9999, type = "webb")}. Effect sizes (Hedges' $g$) were computed and interpreted \cite{cohen_statistical_2013} with \lstinline[language=R]{effectsize::rm_d(method = "av", adjust = TRUE)} \cite{ben-shachar_effectsize_2020}, which used mean aggregation at the team level and the Cumming \cite{cumming_understanding_2013} standardization method. Because we performed many hypothesis tests, we grouped $p$-values into families by research question (RQ1, RQ2, RQ3, and RQ4) and adjusted them with the Benjimini-Hochberg ("BH"; "fdr") method to control the family-wise error rate (i.e. false discovery rate) at $\alpha=.05$  \cite{benjamini_controlling_1995}. 

Finally, qualitative data was used to contextualize and explain each family of results. Rather than constituting a substantive standalone qualitative analysis, this was intended to help triangulate and validate the quantitative findings. As such, the method was very simple and involved one author selecting related quotes from the exit questionnaire responses to represent a diversity of perspectives from participants. To ensure fair representation of participants' perspectives, each participant was quoted exactly once, with selections distributed approximately uniformly across the four research questions. All collaborators read all exit questionnaire responses to confirm that the reported quotes fairly represented expressed sentiments. No formal coding process was performed.

\section{Results}

\subsection{Team Performance} 

\begin{figure}
    \centering
    % \includesvg[width=\linewidth]{visuals//results/perf-plots}
    \includegraphics[width=\linewidth]{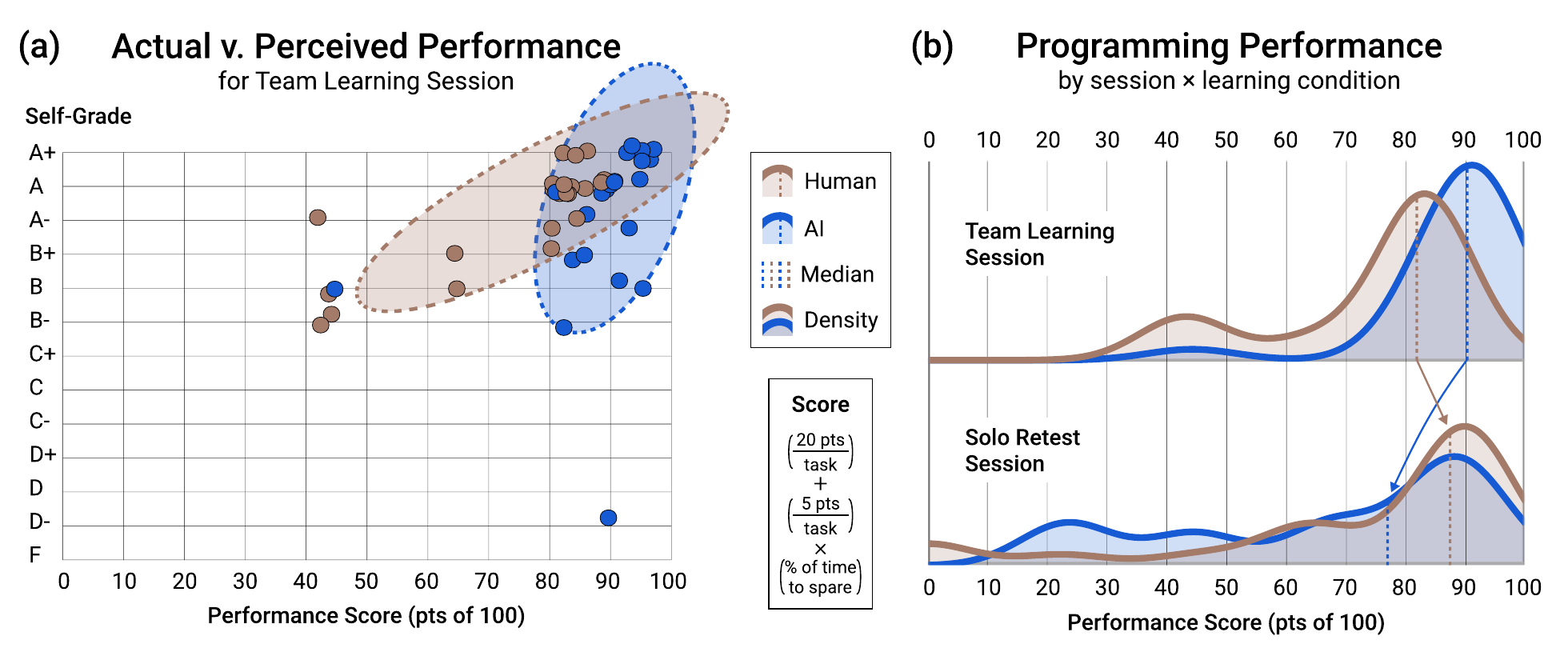}
    \caption{From left to right, (a) shows a scatter plot of actual Session 1 team performance score and self-grade for one's own contribution to Session 1 team efforts and (b) shows density plots of programming performance for team and retest trials by experimental condition.}
    \label{fig:perf-plots}
    \Description{A scatterplot shows a steady positive relationship of performance score from around (40 points, B- grade) to (100 points, A+ grade) for human-human trials and a steep positive relationship from (80 points, B- grade) to (100 points, A+ grade) for human-AI trials of Session 1. Two density plots show the distributions of performance for AI-related and human-related trials on Sessions 1 and 2. AI appears to give a performance advantage for the team learning session and a disadvantage for the solo retest session.}
\end{figure}

\cref{fig:perf-plots}a shows a scatterplot of the actual and perceived performance of human-human and human-AI teams during Session 1, where participants were balancing performance and understanding of their solutions. There is a clear trend for higher scores in the AI condition, with most teams completing all tasks in both conditions but faster with AI. Participants' individual self-grades, which were clearly distinguished from the team's joint performance, still generally followed the teams' actual performance. The family of hypothesis tests for RQ1 is presented in \cref{table:perf}; the family-wise error rate was controlled between these two tests. The performance score was roughly 14 points higher out of 100 when the teammate was Copilot instead of a human, a statistically significant difference ($p_{adj}<.001$). This large \cite{cohen_statistical_2013} effect corresponds to an additional $\frac{7}{10}$ of a task or the same number of tasks but faster by a margin exceeding half of the total time limit. The upper plot in \cref{fig:perf-plots}b shows that both the score distribution and the median score were higher with AI. However, participants' self-grades showed a non-significant ($p>.1$) trend in the opposite direction. In other words, though participants achieved significantly better actual performance with the AI, they did not feel that their individual performance was significantly better in either condition.

\phantomsection\label{para:result-qual-perf}
Participants could tell Copilot gave them a performance advantage, with k17 noting that "AI definitely gets things done a lot faster than humans." They also appreciated its wealth of knowledge; k4 recalled that "there was some Python stuff that both my teammate and I didn't know, but AI could provide that information." This led to conclusions that "if you are trying to get work done, the AI is much more successful" (k6) and usually "goes for an optimal solution" (k22), whereas "humans are wrong much more often" (k6). On the other hand, participants' descriptions of how they collaborated with their teammates helps to explain why they did not take credit for AI enhancement in their self-grades. Participant k12 described a multi-stage process they used with their human teammate, starting with strategies and then comparing alternative options and discussing them prior to implementation. However, with Copilot, they "didn't even think about the task" but "instead, I just copy-and-pasted the PDF and send it to AI." Being more engaged in the human-human condition made k13 feel "like I contributed more to the group." Participant k11 found this communicative process "more stimulating" than messaging Copilot "because you get to discuss what methods you think work and what methods you think don't work." In other words, there is a validation of each teammate's perspective as useful to the team. With Copilot, its superior performance makes the programmer feel as if they are much less needed.

\input{tables/perf-table}

\subsection{Individual Learning} \label{sec:learning-results}
Participants knew they would return one week after Session 1 for a "retention test," i.e., Session 2. Unbeknownst to them, this test consisted of repeating both sets of tasks from the prior session but with no teammate of any kind. This way, their retest performance was a proxy for their learning or understanding of the tasks depending, theoretically, on with which teammate they were first exposed to that particular set of tasks. The lower plot in \cref{fig:perf-plots}(b) shows the distribution of retest scores by previous teammate. Learning with a human showed a trend for higher retest performance. \cref{table:learning} shows the family of hypothesis tests for RQ2; the family-wise error rate was controlled across these four tests. Relative retest performance was worse for the AI condition, a marginally significant difference ($p<.05<p_{adj}<.1$) but absolute retest performance was not significantly different ($p>.1$). The relative constant indicates participants generally performed worse on the retest than they did working in teams, but the zero-inclusive confidence interval means their performance was statistically comparable. However, due to the higher initial performance with AI, the further retest drop by roughly 19 points of 100 in the AI condition constituted a large \cite{cohen_statistical_2013} effect (see also \cref{fig:perf-plots}b).

When considering only the 11 participants indicated as the weaker teammate within each of their human-human teams, the lower mean retest performance was also not statistically significant ($p>.1$). However, when considering only the 11 stronger teammates, the lower retest performance was marginally significant (after $p$-value adjustment; $p<.05<p_{adj}<.1$). Hedges' $g$ indicated this was a small \cite{cohen_statistical_2013} sized effect, at less than four points of 100 on average. The confidence intervals correspond to learning with AI costing stronger teammates between 30 seconds and 7 minutes for an equivalent score on the 20-minute retest compared to learning with the weaker teammate. Notably, though the mean absolute performance difference was larger for weaker teammates, the variability was much higher, with some retesting better on AI learned tasks and others on human learned tasks.

Participants' reflections were also mixed regarding whether the AI or human teammate was better for learning, though they generally seemed to favor the human. Because of its wide knowledge base, Copilot was fast and reliable for helping participants to understand code. Participant k1 found that Copilot "can answer more of your questions than a human can." It was also great for understanding the syntax of Python, which was helpful for many participants who were recruited from Java courses. For example, k15 said, "I felt confident in my ability to do all of the programs in Java, but not so in Python, so using Copilot felt like using training wheels on a bicycle to enhance my initial understanding of Python." However, k15 also found that Copilot often removed the need to think, whereas with the human, "we had to think out our solutions, exercising our brains." Participant k7 echoed the same sentiment, saying, "I think that the AI teammate limited me in my ability to think, whereas that was not the case with the human teammate." This pattern also led k8 to the conclusion that "there was a better understanding of the code that was written between both people." Even though Copilot was probably a better quality learning resource than most human teammates were to each other, the natural ways that they chose to approach the interaction meant that they felt less engaged with the programming concepts in the AI condition.

\input{tables/learning-table}

\subsection{Workload} \label{sec:workload-results}
To compare workload between the two teaming conditions, we explored four of the six dimensions of the NASA TLX: mental demand, temporal demand, effort, and frustration. We excluded physical demand and perceived overall performance as unhelpful for answering RQ3. \cref{fig:wkld-emo-plts}a shows how workload ratings following the team trials differed by condition, with workload generally appearing lower with AI. \cref{table:workload} shows the family of hypothesis tests for RQ3; the family-wise error rate was controlled across these four tests. Mental demand, temporal demand, and effort were all significantly lower with the AI teammate ($p_{adj}<.01$), with large effect sizes \cite{cohen_statistical_2013}. Only frustration, which was low in both conditions, was lower by a non-significant margin ($p>.1$).

\input{tables/tlx-table}

The feedback from the exit questionnaire aligned with the lower workload ratings observed in the AI condition. Participants mentioned how working with Copilot was "leaps and bounds easier" (k18), such that some participants were "not stressed at all" (k21). In general, Copilot's high performance was often mentioned along side a less effortful experience. Building on prior comments about thinking less with Copilot, k9 emphasized how with the human, "you have to use more brainpower." Regarding frustration, which is an emotionally charged aspect of workload, participants did not seem to mention such feelings in either case. This is consistent with the low ratings for both conditions. Participant k3 did mention one circumstance that might be a source of frustration with Copilot, saying "It cannot follow complex logic or understand something that is not on the screen or given to it." In other words, Copilot lacked the "bigger picture" (k17) of implicit needs that humans often assume among each other. Participant k10 mentioned an emotionally salient reason for higher stress in the human condition, instead: "Programming with a human was slightly more stressful and rushed for me because I felt pressure to keep up with the pace at which my teammate was reading through the prompts and coming up with answers." On the other hand, they felt no stress with Copilot because the "fear of human judgment was removed" (k10). Perhaps it is not frustration, but rather temporal demand or some other dimension, which captures this negative experience. Though other results have suggested participants moved through content hastily with Copilot, peer pressure could have also caused weaker teammates to feign understanding before they were really ready to advance, harming their later retest performance.

\subsection{Emotion}

\begin{figure}
    \centering
    % \includesvg[width=\linewidth]{visuals//results/workload-emotion-plots}
    \includegraphics[width=\linewidth]{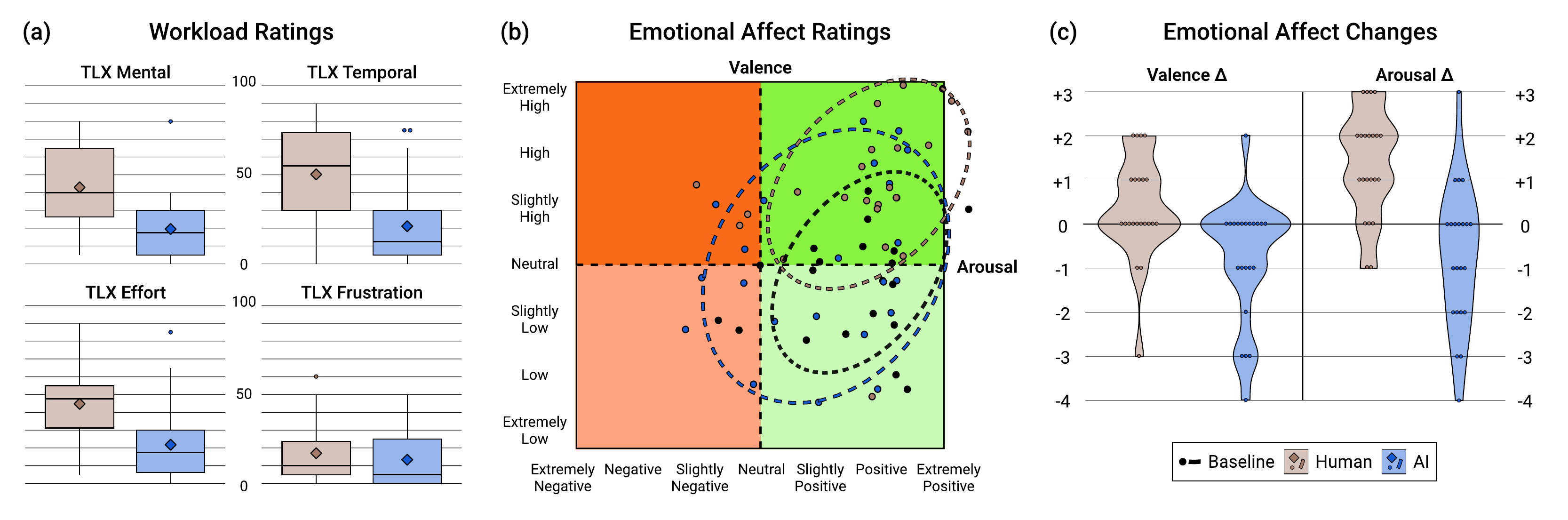}
    \caption{From left to right, (a) shows box plots of four workload sub-scales from the six-item NASA Task Load Index, excluding the physical and performance dimensions, (b) shows a scatterplot of emotional (valence, arousal) pairs taken at three time points during Session 1, the team learning session, and (c) shows violin plots of the emotional impacts (changes) of programming (during Session 1) in the two experimental conditions.}
    \label{fig:wkld-emo-plts}
    \Description{Three plots show that workload is generally lower for AI-related trials than human-related trials, emotion is generally positive in valence and moderate to high in arousal, and emotional impacts appear more negative in both valence and arousal for the AI condition than for the human condition.}
\end{figure}

Emotion ratings were taken three times during Session 1 surrounding team programming. \cref{fig:wkld-emo-plts}b shows that ratings generally reflected positive and aroused moods. Participants were more neutrally aroused at their first measurement, with more positive and aroused emotions after programming with a human and less positive emotions after programming with AI. \cref{fig:wkld-emo-plts}c shows the emotional effects of the programming trials by condition. Note that depending on which ordering participants were assigned, their post-trial mood might have followed the baseline or the opposite condition. The choice of statistical method should account for this dependence. The family of hypothesis tests in \cref{table:emotion} for RQ4 show that the emotional impact of the AI teammate was significantly less positive and less arousing than that of the human teammate; the family-wise error rate was controlled across these two tests. Confidence intervals indicate that this difference was not just a lesser positive effect, but likely a difference in sign in both cases. That is, an AI teammate seems to reduce emotional valence and arousal slightly, while a human teammate seems to increase them both slightly. These large effects \cite{cohen_statistical_2013} arise in a circumstance where participants can easily compare the two conditions. That is, one's mood decreases with AI can be related to following a higher state from the human condition, and vice versa.%; the reverse can also be true. 

% Three-in-one

\input{tables/emo-table}

\phantomsection\label{para:result-qual-emo}
Qualitative data strongly supported the quantitative findings about emotion. As mentioned with the frustration dimension of workload, there does seem to be some crossover between mood and workload. For example, k19 reported that "emotionally speaking, working with a human partner will give you different sense of courage and stress." That is, the human gives you the confidence to work through challenges, but also raises the stakes, heightening emotional arousal and perhaps workload. In terms of valence, participants like k5 found that "working with a teammate felt better because you help each other which gives more fulfillment." Similarly, k14 said that with the human, "I feel like my overall mood improved." Finally, one comment crystallizes how some participants likely struggled to disentangle valence from arousal when providing their ratings. According to k16, working with a human "keeps the mental and social energy up, so I was more happy/excited when coding." They experienced feeling not only "energized" (arousal) but also "happy" (valence) and "excited" (valence and arousal). Another interpretation is that valence and arousal simply tended to move together.

\section{Discussion}
\subsection{RQ1: To what extent does actual and perceived programming performance differ between the human-human and human-AI team paradigms during a learning effort?} \label{sec:discuss-rq1}
Overall, we found that participants performed better with Copilot than in pairs, but this was not reflected by correspondingly high self-grades. Participants did not feel worse about their own performance when using Copilot; they simply did not take personal credit for its help. Qualitative data painted a broader picture, showing that participants felt more useful and valued with human partners, though this outcome wasn't fully reflected in their self-grades. One likely reason for the mismatch between these confidence reports is that Copilot's consistently superior performance made students reluctant to credit themselves for success. Copilot's superior knowledge base and task performance positioned it as the unequivocally stronger teammate, and because its responses were both confident and sycophantic, participants often experienced it as taking over the problem and supplying a perfect solution, leaving little space to see their own contribution. From a Control-Value Theory perspective \cite{pekrun_control-value_2006}, working with Copilot clearly increased outcome value but reduced the programmer's control over that outcome. Copilot's assistance also reduced activity focal value by taking away the meaningful challenge associated with programming.
% Nicholas questions this point because Copilot doing the work reduces cotnrol; however, it does increase outcome value
% JP: I agree and I like your change above.
% have increased students' sense of control by improving immediate task performance.  However, the lower effort required to solve tasks may also reduce the sense that the activity is meaningful or challenging.
In the human team, however, leadership was negotiable and shifting, and shared goals were assumed. Participants knew their teammate was of similar ability and pursuing the same balance between performance and understanding. By contrast, few participants took the time to articulate contextual factors like these to Copilot, contributing to participants finding Copilot to be less collaborative than their human teammate. CVT predicts social object foci (oneself or a compatriot) to evoke positive emotions when either partner can contribute to a shared goal \cite{pekrun_control-value_2024}. For example, being helpful to another inspires satisfaction and pride, while being helped by another evokes gratitude and liking. If the balance of these were to shift, as when partners differ greatly in ability level \cite{hanks_pair_2011} or Copilot outshines the programmer, value can be high, but control is low. That is, a weak partner becomes reliant on the stronger (or Copilot) to obtain either outcome, activity, or social value.
Interventions, such as requiring context specifications or enabling multi-modal communication with AI, could narrow the current gap, but for now, human interaction still supports these collaborative dynamics more effectively than generative AI.

\subsection{RQ2: To what extent does learning retention from a programming effort differ between the human-human and human-AI team paradigms?} \label{sec:discuss-rq2}
There was a trend toward worse learning retention for tasks learned with AI rather than with a human teammate, but these differences were mostly not significant. 
This general trend aligns with recent empirical work showing that AI assistance during learning can impair long-term knowledge retention \cite{barcaui_chatgpt_2025}. Curiously, participants' absolute retest performance actually aligned better to their initial self-grades than to their team performance, suggesting that students may be better judges of their own learning than objective assessments corrupted by AI involvement. Relative to the initially higher performance with AI compared to the human, there was a larger performance drop (marginally significant) for the retest when repeating tasks learned with AI. It might be expected from the theories of Zone of Proximal Development (ZPD; \cite{vygotsky_mind_1978}) or Flow (ZPF; \cite{basawapatna_zones_2013, gardella_performance_2024}), that weaker teammates would learn better by their teammates' pushing them slightly beyond their current ability. Yet, it was instead stronger teammates who had slightly lower (marginally significant) absolute retest performance for the AI tasks. Aside from random chance, one explanation for this is that stronger teammates were acting as teachers in the interaction. Since they had to understand the code well enough to help their weaker counterpart understand it, too, perhaps they engaged with the code more deeply, allowing them to reproduce it more quickly in the retest. This model, called \textit{near-peer teaching} within medical education, does seem to correlate with academic performance for the teaching student, though selection bias is likely present in observational studies \cite{komocsi_academic_2026}.
%Another possibility is that the stronger teammate was the "rate limiter" in the interaction, causing the team to progress  as fast but no faster than their own understanding. 
Another possibility is that stronger teammates served as a "rate limiter", pacing the team according to their own understanding. As noted in \cref{sec:workload-results}, weaker teammates may have hesitated to admit confusion or request a slower pace.
%As noted in \cref{sec:workload-results}, weaker teammates may have feared judgment if they were to admit that the pace was too fast or that current approach was unclear. The confidence intervals in \cref{table:learning} align with this explanation, in that weaker teammates trended toward better learning with the human, but this effect was muddied by some who apparently learned worse. Though perhaps many weaker teammates learned well from their human partners, others may have been able to learn better from Copilot, where they could progress through the learning process at their own pace. 
The confidence intervals in \cref{table:learning} align with this interpretation, in that weaker teammates trended toward better learning with humans, though the effect was inconsistent. While many weaker teammates may have benefited from human explanations, others may have learned more effectively with Copilot, where they could progress at their own pace. 
%A final explanation is also relevant to the inconclusive overall learning result. As was expected, Copilot helped participants complete tasks much more quickly, leaving plenty of time for passive studying at the end of Session 1. In contrast, human-human teams worked through problems carefully together, but risked failing to complete all tasks or finishing with little time to spare. This could be expected to hurt both teammates but especially the weaker teammate, whose natural pace was likely not that taken by the pair. Whereas the stronger teammate helped the team to finish on time based on their understanding, the weaker teammate may have had lingering questions that they did not have time or courage to ask.
Finally, as was expected, Copilot helped participants complete tasks much more quickly thus leaving plenty of time for passive studying at the end of Session 1.  In contrast, human pairs sometimes finished late or with little time to review and consolidate their understanding. This may have disadvantaged weaker teammates in particular, who may have had unresolved questions but lacked time or confidence to raise them.

\subsection{RQ3: To what extent does the workload associated with educational programming differ between the human-human and human-AI team paradigms?} \label{sec:discuss-rq3}
Mental demand, temporal demand, and effort were all significantly lower with Copilot, while frustration was not significantly different. These findings indicate that learning is likely more active in a human-human team, but there is a possibility of covering less material before becoming cognitively fatigued \cite{sweller_cognitive_2011}. The differences in mental demand and effort mirror a prior study comparing solo and AI-assisted paradigms \cite{gardella_performance_2024}. However, that study did not find a conclusive difference in temporal demand. A possible reason for this is that there were many more tasks available than participants would likely have been able to complete in the time limit. The result was that very few participants completed all tasks with any time to spare. By comparison, the task pools here were designed to be completed approximately within the time limit, so many more participants completed them with time to spare. As such, participants here likely felt considerably less rushed in the AI condition, where they were not working up to the time limit. Frustration trended lower in the AI condition, but not significantly so, in line with prior work \cite{gian_luca_scoccia_exploring_2023, gian_luca_scoccia_exploring_2023, gardella_performance_2024}. It seems that participants were not particularly frustrated in either condition. Whereas frustration is a workload state associated with high arousal negative emotion, participants generally reported positive moods throughout the study.

\subsection{RQ4: To what extent does the emotional impact of educational programming as measured by emotional valence and arousal differ between the human-human and human-AI team paradigms?}

Though mood ratings were overall positive in valence and neutral in arousal, the emotional impact of programming was significantly more positive and arousing with the human teammate compared to Copilot. This led to an overall more enjoyable and engaging experience despite lower performance and higher workload. From a Control-Value Theory perspective, these findings suggest that human collaboration increased both the activity value and social value, producing high-arousal positive emotions \cite{pekrun_control-value_2024}. As seen in the \hyperref[para:result-qual-emo]{qualitative emotion results}, social value is the fulfillment from two-way reliance, which maintains control on both sides; the activity itself also becomes more fun and therefore more valuable. Findings also \hyperref[para:result-qual-perf]{supported} that productive team conflict encouraged curiosity at the epistemic incongruity object focus \cite{pekrun_control-value_2024}.
%Compared to the inconclusive learning results, 
This is an exciting finding, suggesting that students are willing to invest more effort in the learning process when working with a human partner than with an AI coding assistant. Research on desirable difficulties shows that moderate challenge can promote learning, whereas tasks that are too easy may reduce meaningful cognitive effort \cite{bjork_memory_1994}. Workload alone therefore provides an incomplete picture of the learning process. While excessive workload can lead to discouragement (or frustration), this was not substantially lower with Copilot, and the more positive emotions observed in the human pairs suggest that the higher workload was productive rather than harmful.
%This is an exciting finding, suggesting students are willing to work harder during the learning process when they work together as opposed to with AI coding assistants. 
%Moderate challenge can promote engagement and learning, whereas tasks that are too easy may reduce meaningful cognitive effort, a principle that underpins research on desirable difficulties \cite{bjork_memory_1994}. Workload alone therefore provides an incomplete picture of the learning process. While excessive workload can lead to discouragement or frustration, this was not substantially lower with Copilot, and the more positive emotions observed in the human condition suggest that the higher workload was productive rather than harmful.
%Workload on its own gives an incomplete picture of the learning process because there is no accepted standard for an exact optimal level for learning. If workload is too low, a student may not be engaging with the aspects of the material that challenge them to grow from their current state of knowledge or ability. The hazard of excessive workload is typically discouragement (or frustration), which was not considerably lower with Copilot. Given that emotions were so positive with the human, it seems that the higher workload was not excessive, but likely productive or at least harmless. 
Arousal tells a similar story, with the mean difference of two Likert points out of seven pointing to a heightened alertness within the true social interaction. Given that optimal arousal is important for human performance \cite{akizuki_measurement_2015, sharpe_sustained_2025}, it is likely that the participants were performing better as individuals in the human team even as Copilot artificially inflated their AI-assisted scores. This was also reflected by the self-grade findings discussed in \cref{sec:discuss-rq1,sec:discuss-rq2}.
\subsection{Implications \& Recommendations} \label{sec:implications}

We strongly recommend that educators look for opportunities to integrate human-human pair programming into their programming courses. The strong emotional benefits amidst lower performance and workload suggest that students will work harder more happily with their peers than with AI. There is a clear benefit and appeal for students to use Copilot on their coursework. It will make programming easier and faster for simple problems, potentially without the major learning consequences that many fear. Going beyond the conditions of this study, however, it is easy to imagine how human-human pair programming could surpass AI-assisted programming. Most participants here had plenty of spare time to study passively with Copilot, making efficient use of a fixed, 20-minute period for that purpose. The slower and more active studying process of the human-human condition either left participants with better understanding of less material (i.e., tasks completed) or a better but incomplete understanding of a similar amount of material. What is not clear is how additional time might be used in a non-experimental setting. With Copilot, participants rarely needed additional time to further study the tasks. 
%In an extreme case, one participant was observed becoming so bored by studying the solutions that he began browsing a shopping website for trading cards to fill the time. 
With the human teammate, however, participants were so energized and engaged that they might reasonably have taken extra time if they were allowed. 
%For example, one pair excitedly asked after their timer expired whether they were allowed to Google a Python question when they left (which, of course, they were not allowed to do). Had the option been given, participants' choices would probably reflect their time availability and desire for compensation more than their will to learn. Though, in a real world scenario, it is reasonable to guess that students might invest more time into a satisfying human social interaction, thus closing the performance gap associated with Copilot's speed.

Similar to prior work \cite{prather_widening_2024,vadaparty_cs1-llm_2024}, our results indicate that Copilot can still be used as an effective educational tool and should not be dismissed completely. With the observed performance benefit and inconclusive learning effect, Copilot seems to offer very accessible and low-stakes information retrieval for students desiring to learn. Prior work has shown evidence that human-human teams can be augmented with AI for a best-of-both worlds scenario \cite{lyu_will_2025}. It seems plausible from our findings here that having Copilot or another AI-based tool to fill knowledge gaps would have been helpful to some of our human-human teams. However, the loss of control and autonomy reported in prior work~\cite{berger_my_2025} was also replicated here and makes introducing AI into human-human teams hazardous. While power dynamics already govern human pair programming \cite{deitrick_how_2016}, adding an omniscient oracle risks an automatic deferral away from student control. One possible solution would be to encourage pairs of students to voluntarily modify the behavior of their AI tools to align with their learning goals via intentional use of system guardrails. 
%For example, Copilot accepts instruction files where users can write about how they want Copilot to behave and what assumptions it should make. Students might choose to request simple explanations, restrict code snippets in chats, or ask for pause points to remind the pair to discuss the problem verbally without Copilot before continuing with its help. 
Another possible feature, which Copilot does not offer, is user-defined rate limiting. Perhaps if students could view Copilot's help as more scarce, they might take better advantage of their human teammate in the average case, only turning to Copilot when stuck. Prior work has shown that students who turn to AI later in the problem-solving process tend to perform better \cite{margulieux_self-regulation_2024}. This could also help with students being annoyed by Copilot's proactive, sometimes too eager, autocompletion suggestions observed in prior work~\cite{prather_its_2024}. Overall, while there is no reason that the benefits of both paradigms cannot theoretically be combined, extra care needs to be taken to ensure that the social benefits of human interaction are not disrupted by adding Copilot into the mix--- prior work has found that easy access to generative AI models is already eroding social interactions in introductory programming classes~\cite{hou_all_2025}.

\section{Threats to Validity}

Regarding internal validity, reliable inference is very challenging with a small sample and the multi-way clustering inherent to this area of inquiry. We defend our statistical method as a best-in-class approach to handling our data but acknowledge that spurious results (Type II errors) are possible. We used $p$-value adjustments within each of four families (one per RQ) to mollify risks to our conclusions. Regarding Type I error, we reported confidence intervals and effect sizes to contextualize our results. With respect to endogeneity, we designed our study with careful counterbalancing procedures in mind and made our two task pools as similar as possible. To ensure no major confounding effects remained from ordering or task pool differences, we also used Wald tests to challenge the parsimony of our regressions. A final concern is the dependence structure of emotion measures. The within-subjects design means that participants' changes in emotion were possibly influenced in magnitude based on the ordering of their trials. Counterbalancing should have mitigated any bias to the conclusion of RQ4, but the absolute magnitude of emotion changes might show higher uncertainty based on the impossibility of resetting one's emotion back to baseline. As such, the absolute changes in mood ratings should be interpreted with caution.

As far as external validity, this work was conducted with a small group of highly organized volunteers at a selective research university in the eastern United States, so attempts to generalize our findings should consider this context. The pairing of participants was also done by convenience and may not reflect an intentional pairing strategy performed in an instructional setting \cite{hanks_pair_2011}. As discussed in \cref{sec:implications}, we suspect that participants might have behaved differently in the two experimental conditions had they been offered additional time in a naturalistic setting. Ultimately, real students will not spend a fixed amount of time learning regardless of whether they collaborate with a human or an AI assistant. More likely, an internal risk-reward balance will determine how much time they invest \cite{son_metacognitive_2006}. Further research should explore what factors determine how much time a student is willing to invest into educational collaborative programming when allowed to choose for themself. Finally, it is challenging to balance design consistency with the realistic complexity of real-world programming tasks. Copilot was known to be reliable at solving the tasks selected from HumanEval, but it remains to be seen whether messier or more poorly specified tasks might have been performed or learned better through a human-human paradigm. Additionally, it is possible that for such tasks, the human might need to be more involved in helping Copilot, such that differences observed here might fade or change.

\section{Conclusion}

This mixed-methods laboratory study used data from an experiment with 11 pairs of novice and intermediate programmers to compare performance, learning, workload, and emotion between human-human and human-AI team paradigms. Prior work suggested that an AI teammate would improve objective performance and reduce workload compared to a human teammate, but it was not clear how self-perceived performance, learning, or emotion would differ. We used linear regression and Wild Cluster Bootstrapping to perform within-subjects comparisons between the two experimental conditions, which were administered in a counterbalanced design. A qualitative questionnaire helped to explain the quantitative results. We found significantly higher performance and lower mental demand, temporal demand, and effort in the AI condition but mostly inconclusive effects on self-perceived performance, frustration, and learning. There was some very limited evidence for worse learning in the AI condition, especially for participants who were the stronger member of their human-human team. The emotional effect of programming was significantly more positive and arousing in the human condition. We conclude that students will likely work harder and more happily in educational programming contexts with human teammates than with GitHub Copilot. We recommend that human-human pair programming be reconsidered for its power as an educational tool, and recommend that students maintain autonomy and social activity if they choose to augment their human-human teams with AI assistance.

\begin{acks}
We thank Jennifer Burman and the UVA Link Lab for supporting the AI Coding Workshop series where participants were trained. This work was supported by the University of Virginia Distinguished Fellowship, NSF National Research Traineeship for Cyber-Physical Systems (1829004), Graduate Research Fellowship Program for Computer and Information Science and Engineering (1842490), and the Virginia Commonwealth Cyber Initiative Central Virginia Node (GR102900). 
\end{acks}

\bibliographystyle{ACM-Reference-Format}
\bibliography{zotero}

\end{document}

%% file: tables/perf-table.tex
\begin{table}

\caption{Regression Coefficients of Treatment on Performance. Bolded rows are statistically significant at $\alpha = .05$. CRSE stands for Cluster Robust Standard Error (Wild Cluster Bootstrapped), $p_{adj}$ for Benjimimi-Hochberg~\cite{benjamini_controlling_1995} corrected p-value for multiple-testing, and g is Hedges' g (effect size).}

\centering
\begin{tabular}[t]{llllllll}

\toprule
\textbf{Variable} & \textbf{\textit{B}} & \textbf{\textit{CRSE}}  & \multicolumn{2}{c}{\textbf{95 \% CI}} & \textbf{\textit{p}} & \boldsymbol{$p_{adj}$} & \textbf{ \textit{g}}   \\
\cmidrule{4-5}
 &  &  & \makecell{\textbf{\textit{LL}}} & \makecell{\textbf{\textit{UL}}} &  & \\
\midrule

\multicolumn{8}{c}{\textit{Actual Team Performance (Session 1)}}\\
\addlinespace
Constant & 74.30 & 5.08 & 62.67 & 84.05 &  & \\
\textbf{Teammate = AI} & \textbf{14.09} & \textbf{3.96} & \textbf{6.46} & \textbf{22.89} & \textbf{<.001} & \textbf{<.001} & \textbf{.99}\\
\midrule

\multicolumn{8}{c}{\textit{Self-Grade for Team Performance (Session 1)}}\\
\addlinespace
Constant & 11.18 (A-) & 0.43 & 10.20 (B+) & 12.06 (A) &  & \\
Teammate = AI & -0.23 & 0.27 & -0.80 & 0.34 & .435 & .435 & -0.13\\

\bottomrule
%\multicolumn{8}{l}{\footnotesize \textsuperscript{a} Cluster Robust Standard Error (Wild Cluster Bootstrapped) } \\
%\multicolumn{8}{l}{\footnotesize \textsuperscript{b} Benjimini-Hochberg \cite{benjamini_controlling_1995} correction for multiple-testing} \\
%\multicolumn{8}{l}{\footnotesize \textsuperscript{c} Hedges' \textit{g} (effect size)} \\
\end{tabular}
\label{table:perf}
\end{table}

%% file: tables/learning-table.tex
\begin{table}

\caption{Regression Coefficients of Treatment on Retest Performance (Learning). Bolded rows are statistically significant at $\alpha = .05$. CRSE stands for Cluster Robust Standard Error (Wild Cluster Bootstrapped), $p_{adj}$ for Benjimimi-Hochberg~\cite{benjamini_controlling_1995} corrected p-value for multiple-testing, and g is Hedges' g (effect size).}

\centering
\begin{tabular}[t]{llllllll}

\toprule
\textbf{Variable} & \textbf{\textit{B}} & \textbf{\textit{CRSE}}  & \multicolumn{2}{c}{\textbf{95 \% CI}} & \textbf{\textit{p}} & \boldsymbol{$p_{adj}$} & \textbf{ \textit{g}}   \\
\cmidrule{4-5}
 &  &  & \makecell{\textbf{\textit{LL}}} & \makecell{\textbf{\textit{UL}}} &  & \\
\midrule

% \addlinespace
\multicolumn{8}{c}{\textit{Retest Performance Change (Session 2 - Session 1)}}\\
\addlinespace
Constant & -2.65 & 6.03 & -15.74 & 10.34 &  & \\
Session 1 Teammate = AI & -18.88 & 6.33 & -32.87 & -5.01 & \textbf{.015} & .054 & -1.13\\
\midrule

\multicolumn{8}{c}{\textit{Absolute Retest Performance (Session 2)}}\\
\addlinespace
Constant & 71.65 & 5.58 & 59.40 & 83.81 &  & \\
Session 1 Teammate = AI & -4.79 & 5.58 & -16.86 & 7.46 & .397 & .529 & -0.27\\
\midrule

\multicolumn{8}{c}{\textit{Absolute Retest Performance for Weaker Teammates Only}}\\
\addlinespace
Constant & 55.01 & 10.42 & 31.87 & 77.38 &  &  & \\
Session 1 Teammate = AI & -6.18 & 11.52 & -30.46 & 18.96 & .606 & .606 & -0.20\\
\midrule

\multicolumn{8}{c}{\textit{Absolute Retest Performance for Stronger Teammates Only}}\\
\addlinespace
Constant & 88.29 & 2.74 & 82.35 & 92.39 &  &  & \\
Session 1 Teammate = AI & -3.40 & 1.57 & -6.88 & -0.40 & \textbf{.027} & .054 & -0.28\\

\bottomrule
%\multicolumn{8}{l}{\footnotesize \textsuperscript{a} Cluster Robust Standard Error (Wild Cluster Bootstrapped) } \\
%\multicolumn{8}{l}{\footnotesize \textsuperscript{b} Benjimini-Hochberg \cite{benjamini_controlling_1995} correction for multiple-testing} \\
%\multicolumn{8}{l}{\footnotesize \textsuperscript{c} Hedges' \textit{g} (effect size)} \\
\end{tabular}
\label{table:learning}
\end{table}

%% file: tables/tlx-table.tex
\begin{table}

\caption{Regression Coefficients of Treatment on Subjective Workload. Four different regression models were fitted, with uncertainty quantified by refined Wild Cluster Bootstrapping. Note that the TLX Physical and TLX Performance dimensions were excluded due to lack of relevance to our research questions. Self-grade is a better measure of self-perceived performance as discussed in \autoref{sec:datacollection}. Correcting \textit{p}-values for multiple testing would not change the conclusions here, even with the very conservative Bonferroni adjustment. CRSE stands for Cluster Robust Standard Error (Wild Cluster Bootstrapped), $p_{adj}$ for Benjimimi-Hochberg~\cite{benjamini_controlling_1995} corrected p-value for multiple-testing, and g is Hedges' g (effect size).}

\centering
\begin{tabular}[t]{llllllll}

\toprule
\textbf{Variable} & \textbf{\textit{B}} & \textbf{\textit{CRSE}}  & \multicolumn{2}{c}{\textbf{95 \% CI}} & \textbf{\textit{p}} & \boldsymbol{$p_{adj}$} & \textbf{ \textit{g}}  \\
\cmidrule{4-5}
 &  &  & \makecell{\textbf{\textit{LL}}} & \makecell{\textbf{\textit{UL}}} &  & \\
\midrule

\multicolumn{8}{c}{\textit{TLX Mental Demand}}\\
% \textbf{\hspace{1em}Variable} & \textbf{\textit{B}} & \textbf{\textit{CRSE}} & \textbf{\textit{LL}} & \textbf{\textit{UL}} & \textbf{\textit{p}} & \textbf{\textit{g}}\\
% \midrule
Constant & 42.95 & 5.74 & 30.52 & 55.59 & & & \\
\textbf{Teammate = AI} & \textbf{-23.41} & \textbf{5.75} & \textbf{-36.42} & \textbf{-11.77} & \textbf{.002} & \textbf{.003} & \textbf{-1.22}\\
\midrule
\multicolumn{8}{c}{\textit{TLX Temporal Demand}}\\
% \textbf{\hspace{1em}Variable} & \textbf{\textit{B}} & \textbf{\textit{CRSE}} & \textbf{\textit{LL}} & \textbf{\textit{UL}} & \textbf{\textit{p}} & \textbf{\textit{g}}\\
% \midrule
Constant & 50.23 & 7.17 & 34.54 & 65.97 & & & \\
\textbf{Teammate = AI} & \textbf{-29.09} & \textbf{6.63} & \textbf{-43.68} & \textbf{-14.53} & \textbf{.001} & \textbf{.002} & \textbf{-1.21}\\
\midrule
\multicolumn{8}{c}{\textit{TLX Effort}}\\
% \textbf{\hspace{1em}Variable} & \textbf{\textit{B}} & \textbf{\textit{CRSE}} & \textbf{\textit{LL}} & \textbf{\textit{UL}} & \textbf{\textit{p}} & \textbf{\textit{g}}\\
% \midrule
Constant & 44.77 & 4.89 & 34.03 & 55.68 & & & \\
\textbf{Teammate = AI} & \textbf{-22.95} & \textbf{4.35} & \textbf{-32.85} & \textbf{-13.37} & \textbf{.001} & \textbf{.002} & \textbf{-1.28}\\
\midrule
\multicolumn{8}{c}{\textit{TLX Frustration}}\\
% \textbf{\hspace{1em}Variable} & \textbf{\textit{B}} & \textbf{\textit{CRSE}} & \textbf{\textit{LL}} & \textbf{\textit{UL}} & \textbf{\textit{p}} & \textbf{\textit{g}}\\
% \midrule
Constant & 17.05 & 4.55 & 7.16 & 27.14 & & &\\
Teammate = AI & -3.64 & 4.84 & -14.57 & 6.58 & .470 & .470 & -0.24\\

\bottomrule
%\multicolumn{8}{l}{\footnotesize \textsuperscript{a} Cluster Robust Standard Error (Wild Cluster Bootstrapped) } \\
%\multicolumn{8}{l}{\footnotesize \textsuperscript{b} Benjimini-Hochberg \cite{benjamini_controlling_1995} correction for multiple-testing} \\
%\multicolumn{8}{l}{\footnotesize \textsuperscript{c} Hedges' \textit{g} (effect size)} \\
\end{tabular}
\label{table:workload}
\end{table}

%% file: tables/emo-table.tex
\begin{table}

\caption{Regression Coefficients of Treatment on Emotion. Dependent measures were rating changes over a trial (after - before) to represent the emotional impact of the trial. CRSE stands for Cluster Robust Standard Error (Wild Cluster Bootstrapped), $p_{adj}$ for Benjimimi-Hochberg~\cite{benjamini_controlling_1995} corrected p-value for multiple-testing, and g is Hedges' g (effect size).}

\centering
\begin{tabular}[t]{llllllll}

\toprule
\textbf{Variable} & \textbf{\textit{B}} & \textbf{\textit{CRSE}} & \multicolumn{2}{c}{\textbf{95 \% CI}} & \textbf{\textit{p}} & \boldsymbol{$p_{adj}$} & \textbf{ \textit{g}} \\
\cmidrule{4-5}
 &  &  & \makecell{\textbf{\textit{LL}}} & \makecell{\textbf{\textit{UL}}} &  & \\
\midrule

\multicolumn{8}{c}{\textit{Emotional Valence Change}}\\
\addlinespace
Constant & 0.36 & 0.14 & 0.07 & 0.64 &  & & \\
\textbf{Teammate = AI} & \textbf{-1.18} & \textbf{0.31} & \textbf{-1.84} & \textbf{-0.52} & \textbf{.001} & \textbf{.002} & \textbf{-1.61} \\
\midrule

\multicolumn{8}{c}{\textit{Emotional Arousal Change}}\\
\addlinespace
Constant & 1.36 & 0.27 & 0.77 & 1.95 &  &  & \\
\textbf{Teammate = AI} & \textbf{-2.09} & \textbf{0.58} & \textbf{-3.37} & \textbf{-0.83} & \textbf{.004} & \textbf{.004} & \textbf{-1.74 } \\

\bottomrule
%\multicolumn{8}{l}{\footnotesize \textsuperscript{a} Cluster Robust Standard Error (Wild Cluster Bootstrapped) } \\
%\multicolumn{8}{l}{\footnotesize \textsuperscript{b} Benjimini-Hochberg \cite{benjamini_controlling_1995} correction for multiple-testing} \\
%\multicolumn{8}{l}{\footnotesize \textsuperscript{c} Hedges' \textit{g} (effect size)} \\
\end{tabular}
\label{table:emotion}
\end{table}